# THE ECLIPSE OF 21 JUNE 1629 IN BEIJING IN THE CONTEXT OF THE REFORM OF THE CHINESE CALENDAR

**Sperello di Serego Alighieri**
*INAF – Osservatorio Astrofisico di Arcetri, Largo Enrico Fermi 5,
50125 Firenze, Italy.*
E-mail: sperello@arcetri.astro.it

**and**

**Elisabetta Corsi**
*Sapienza Università di Roma, Piazzale Aldo Moro 5,
00185 Roma, Italy.*
E-mail: elisabetta.corsi@uniroma1.it

**Abstract:** This paper examines the predictions made by Chinese, Muslim and Jesuit astronomers of the eclipse of 21 June 1629 in Beijing, allegedly the event that determined Emperor Chongzhen's resolution to reform the calendar using the Western method. In order to establish the accuracy of these predictions, as reported at the time by the Chinese scholar and convert Xu Guangqi, we have compared them with an accurate reconstruction of the eclipse made at NASA. In contrast with current opinions, we argue that the prediction made by the Jesuits was indeed the most accurate. It was in fact instrumental in dissipating Chongzhen's doubts about the need to entrust Jesuit missionaries serving at the Chinese court with the task of reforming the calendar, leading to the first important scientific collaboration between Europe and China.

## 1 INTRODUCTION

Eclipses are inauspicious events in Chinese traditional worldview, bearers of major disasters. For centuries astronomers serving at the Chinese Court—from the Tang Dynasty on, there were also foreign astronomers, firstly Brahmans then Arabs and lately Europeans—strove to make the best eclipse forecasts, lest the Emperor, who was believed to be the *tianzi* 天子, the Son of Heaven and a link between Heaven and Earth, might cause havoc among his subjects by acting in an unworthy way under the evil spell cast by the celestial event.

Jesuit missionaries soon realized the high status that astronomy and eclipse prediction enjoyed among the ruling elite of China and used their knowledge in the field to compete with court astronomers, so that they succeeded in persuading the Chongzhen Emperor that they were more qualified than Chinese and Muslim court astronomers to perform the long needed task of amending the calendar. As Catherine Jami (1995: 174) pointed out, "Precision in the prediction of celestial phenomena might be a matter of life or death for the officials in charge. Astronomy and time-keeping both had heavy political implications, and their importance in that respect should be kept in mind."

The Jesuit involvement with the Astronomical Bureau (*Qintianjian* 欽天監) in Beijing, an Imperial establishment whose tasks were essentially those of providing the official calendar to the Empire, and making astronomical predictions and calculations, may not be fully appreciated unless attention is paid to the quality of the Jesuit scientific and technical expertise, a condition that enabled them to retain a leading position in the Astronomical Bureau for a century and a half, starting with Adam Schall von Bell's appointment in 1644, at the establishment of the Qing Dynasty. In order to do so, we begin by outlining the context in which this expertise was acquired.

## 2 CONTEXT

Indeed, a major calendrical reform had already taken place in Europe about fifty years earlier and the Jesuits had been instrumental in carrying it out successfully. The year 1582 is when the so-called Gregorian reform of the calendar was implemented in Europe under Pope Gregory XIII (1502–1585). As it is known, the need for the reform rose from the fact that discrepancies between the tabulated Spring equinox and the observed one had been increasing to a lapse of ten days in Pope Gregory's time, causing serious problems to the date of Easter (Heilbron, 1999; see especially 24–46, 144–175).

The year 1582 is also when the Jesuit missionary Matteo Ricci (1552–1610) arrived in China to proselytize (Brockey, 2007; Fraanke, 1976; Hsia, 2010). As it is known, he managed to reach the Imperial Court in Beijing thanks to the knowledge of mathematics and natural philosophy he had acquired during the years spent at the Collegio Romano. Established in 1551, the Collegio Romano was soon to become one of the most reputed Jesuit high education institutions in Europe and a model for similar establishments in the colonial world (Bernard, 1935; Corsi, 2012; Grendler, 2002). Thanks to the *Ratio atque Institutio Studiorum Societatis Iesu* (1599, see Hinz et al., 2004), astronomy, cosmology, as well as a cluster of sub-disciplines that fell under the denomination of 'mixed mathematics', were formally included in the curriculum of studies in major Jesuit colleges. As Jesuits' reputation as educators of European elite grew, the principles and regulations of the *Ratio* began to affect the production and circulation of knowledge in high education institutions of the Catholic world, to the extent that even their antagonists could not but acknowledge their pedagogical mastery.

Soon after his return to Rome from Coimbra in 1561, Christopher Clavius (1538–1612) had been teaching mathematics at the Collegio Romano, in what became known as Clavius' Academy. At the Academy, Clavius taught more advanced courses than the public ones, but they went on rather informally until about 1594, when Clavius' requests to the superiors for its official recognition were finally accepted. Clavius gave two reasons for the establishment of an Academy of Mathematics at the Collegio Romano. Firstly, if Jesuit educational institutions were to compete with public universities, especially in Protestant Europe, they should be able to offer instruction of the highest level, not only in philosophy and theology, but also on a broad range of secular disciplines of a more technical and practical nature. Secondly, Jesuits who were to travel to mission stations should gain a scientific and technical expertise in order to fulfill the demands of their daily lives overseas (Baldini, 2002; Romano, 1999). This point is of particular interest for the reasons that will be provided further on in this paper.

By the time Ricci entered the Society, the Collegio Romano was running its third programme of public studies, one in which arithmetic was no longer included in the first year of the philosophy course but had been transferred to the second year:

> [In the second year of philosophy]: the first four books of Euclid during four months approximately, Practical Arithmetic one month and a half, Sphere two months and a half, Geography two months, and, during the remainder of the year, Books 5 and 6 of Euclid.
>
> [In the third year of philosophy]: Astrolabe two months, Theory of the Planets four months, Perspective three months, during the remainder of the time Clocks and ecclesiastical Computation (*Monumenta Paedagogica*, 1901, cit. in Bernard, 1935: 30).

From the above description, we can observe that the instruction in 'mixed mathematics' was quite elementary. Nonetheless, the more talented students could request their teachers for additional private lectures or they could attend the Academy of Mathematics. Ricci too attended lectures held by Christopher Clavius, although only after 1594 did the *catalogi* of the Collegio Romano "… begin to specify a distinct group of *mathematicians* (the official title of those attending the Academy)." (Baldini, 2002: 52).

A host of other Jesuits such as Niccolò Longobardi (1565–1654), Diego de Pantoja (1571–1618), Sabatino de Ursis (1575–1620), Nicolas Trigault (1577–1628), Johannes Schreck (1576–1630), Giulio Aleni (1582–1649), Wenceslas Kirwitzer (1588–1626), Adam Schall von Bell (1591–1666), Giacomo Rho (1593–1638), Martino Martini (1614–1661), and Ferdinand Verbiest (1623–1688) followed in the footsteps of Ricci, bringing to the Chinese court further astronomical knowledge and contributing to solve the problems of the Chinese calendar. Among them, only Sabatino de Ursis and Giulio Aleni had attended the Academy.

De Ursis was there from 1600 to 1601 and Aleni from 1606 to 1608. It is worth noting that De Ursis was in Beijing from 1607, with Ricci, until he was expelled in 1617. During this period he actively contributed to the calendar reform and composed a few treatises in Chinese on mechanics and hydraulics. Since his stay at the Academy was quite short, we have to assume that he may have acquired most of his scientific knowledge during the years spent at Naples as a novice and later on in Macao, where he returned after his expulsion from Beijing (Gatto, 1994; 1995: 283–294). Indeed, Clavius resided at the Collegio di Napoli for a year from 1595 to establish an Academy of Mathematics there; his pupil from the Collegio Romano, Giovanni Giacomo Staserio (1565–1635) was to take the chair in Mathematics at Naples and it is from him that De Ursis received his early scientific training (Baldini, 2000: p. 93, n. 94, p. 94, n. 98).

Aleni's scientific expertise seems to have been even more accomplished, as he had studied at the prestigious Collegio di Parma before being admitted to the Collegio Romano. He left Lisbon for Goa in 1609 and once he reached the Portuguese enclave he observed a lunar eclipse about which he reported to Magini in a letter dated January 1611 (Baldini, 2002: 97, n. 106). As we shall see in the following paragraphs, the astronomical expertise acquired by Jesuit missionaries has been crucial for their penetration in the Chinese world, particularly at the highest levels of the establishment.

Let us now consider the entangled issues related to the attempts of reforming the Chinese luni-solar calendar made by the Jesuit missionaries serving at the Chinese court. Just as Clavius was among those who persuaded Pope Gregory XIII that the time had come for a readjustment of the calendar which, among other things, would provide a correct date for Easter, De Ursis and his confreres used eclipse prediction to win the Emperor's trust over their rival Chinese and Muslim astronomers, and obtain his consent to implement the reform. In short, the problems of any calendar originate from the fact that the year is not made of an integer number of days, but it is 365.24219 days long.

The Julian calendar, promulgated by Julius Caesar in 46 BCE with the advice of Sosigenes of Alexandria, assumed the year to be 365.25 days long, which is 11.25 minutes wrong; so the Julian calendar went off by one day after 128 years. The Gregorian reform instead assumed the year to be 365.2425 days long, which is wrong by only 27 seconds, so it will take more than 3,200 years for the Gregorian calendar to go wrong by one day. The problems of the Chinese calendar were even more serious, because, being a luni-solar calendar, it had to stay in phase not just with the rotation of the Earth around the Sun (the year), but also with the rotation of the Moon around the Earth, i.e. with the lunar month, which is not made of an integer number of days (it is 29.53059 days long), nor the year is made of an integer number of lunar months (it is 12.3683 lunar months long). In the sixteenth century the Ming Emperors had problems with the calendar, which turned into a mismatch between celestial phenomena and the calendar. Attempts to reform it had been unsuccessful, opening the way to Jesuit astronomers (Chu, 2007; D'Elia, 1947).

## 3 ANALYSIS OF THE ECLIPSE OF 21 JUNE 1629

In September 1629 the Chongzhen Emperor (1611–1644), the last monarch of the Ming Dynasty, assigned the task to reform the Chinese calendar with the Western method to Xu Guangqi (1562–1633), a civil servant belonging to the Confucian elite, who had converted to Christianity in 1603 thanks to his friendship with Matteo Ricci. Reports state that the crucial event for this decision was a solar eclipse, which happened on 21 June 1629. The eclipse was partial in Beijing, and had been forecaste both by the Chinese astronomers, who made use of the traditional *Datongli* 大統曆 method; as well as by Muslim astronomers, whose computational method was known as *Huihuili* 回回曆. Jesuit astronomers too made their own prediction by means of the Western method of computation. Given the fact that a couple of months later the Chongzhen Emperor decided that the Calendar should be reformed using the Western method, one would assume that the Jesuits prediction was more accurate than the Chinese and Muslims ones, and therefore there should be no reason to further investigate the case.

Nonetheless, an analysis of recent scholarship about this eclipse and its forecasts shows contradictions that make it worth delving more thoroughly into the case. Let us proceed with a review of the major studies concerning the eclipse forecasts. One of the first authors to be taken into consideration is Agustín Udías, who writes that the prediction was made by Johann Schreck (known as Terrentius) by means of the Western method, and that it was the only correct one (Udìas, 1994). This assertion is supported also by Daniel J. Boorstin (1983: 62) who writes that:

> The Imperial Astronomers predicted that the eclipse would occur at 10:30 and would last for two hours. The Jesuits forecast that the eclipse would not come until 11:30 and would last only two minutes. On the crucial day, as 10:30 came and went the sun shone in full brilliance. The Imperial Astronomers were wrong, but were the Jesuits right? Then, just at 11:30, the eclipse began and lasted for a brief two minutes, as the Jesuits had predicted.

A similar statement had also been made by D'Elia (1947).

On the other hand Lü (2007) writes that "… the truth is totally different …" and, in order to prove his argument, he provides a comparison table (Lü, 2007: Table 1), showing that

> … the error of the maximum phase and last contact of the Western method is much larger than that of the *Datong li*, whereas the error of the magnitude and first contact of the Western method is smaller than that of the *Datong li*. Checked against the observation results, the error of the magnitude, maximum phase and last contact of the Western method is also much larger than that of the *Datong li,* while only the error of the first contact is slightly smaller than that of the *Datong li*. In sum the results of the prediction provided by the *Datong li* were very bad, but those of the Western methods were even worse.

Unfortunately, Lü's argument leaves room for doubts, because the data contained in his Table 1 show some anomalies: the duration of the eclipse is not reported, the time of the maximum phase in the rows about Western method and observation is not halfway between the first and last contacts, as it should be in an eclipse, and it is not clear where the data for the row with the *Theoretical Results* come from.

In order to shed light on this important event, we checked the original reports about the predictions and the observation of the eclipse of 21 June 1629 gathered by Xu Guangqi 徐光啟, in *Zhili yuanqi* 治曆緣起 *The Beginning of the Calendar Reform* (1645), and we compared them with the accurate reconstruction of that solar eclipse at Beijing Ancient Observatory, made using the Eclipse Predictions by Fred Espenak and Chris O'Byrne (NASA's GSFC) (https://eclipse.gsfc.nasa.gov/JSEX/JSEX-index.html). The data concerning the predictions made using the traditional method (*Datong li*), the Muslim method (*Huihui li*), and the Jesuit method (Western), and the eclipse observations as reported in the *Zhili yuanqi*, and the NASA reconstruction (NASA) are shown here in Table 1. Times are given in hours and minutes of local apparent time.

Since the NASA reconstruction gives mean local times for the time zone, we have corrected them taking into account both the difference in longitude between the Beijing Ancient Observatory and the relevant time zone ($-14^m 16^s$) and the equation of time for 21 June ($1^m 20^s$). We have assumed that Xu Guangqi has used the Western division of the day in 96 units (刻 *ke*) of 15 minutes each (cf. Jami, 1995), instead of the classical Chinese division in 100 *ke* of 14.4 minutes, which is assumed in LL07. The *Zhili yuanqi* also reports that for the eclipse of 21 June 1629 the units smaller than one hour are rendered as *suan-wai* ("outside the count"), i.e. they are added at the end of the larger units (see also Chu and Shi, 2014). The magnitude is the fraction of the Sun's diameter obscured by the Moon at maximum; the solar diameter is divided into 10 *fen*, each divided into 60 *miao* (Stephenson and Fatoohi, 1995), while other authors have assumed that one *fen* is divided into 100 *miao* (e.g. Lü, 2007). The duration shown in the first four rows of Table 1 is that given in the *Zhili yuanqi*. The last three rows of Table 1 give the difference between the NASA reconstruction and the forecasts made with the Datong, Huihui and Western methods.

A quick look at the data reported in Table 1 is all one needs to note that, as a matter of fact, forecasts were not accurate, the timing errors often being larger than 15 minutes, although they do not seem much worse than the errors for eclipse forecasts made by Jesuit astronomers between 1644 and 1750 for which the errors go up to 41 minutes and are often above 15 minutes (Stephenson and Fatoohi, 1995). It is worth highlighting that taking into the correct account the *suanwai* allowed us to solve the problem of the maximum phase not being halfway between the first and the last contacts, as shown in Lü (2007: Table 1) for the prediction with the Western method and for the observations.

The report of the eclipse of 21 June 1629 in the *Zhili yuanqi* also mentions an orientation, which is South-West for the first contact, South for the maximum, and South-East for the last contact; this is the same for the forecasts with the three different methods and for the observation. Clearly this orientation refers to the position of the Sun in the sky, since the Sun moves from East to West. We think that it refers to the position of the Moon on the Sun's surface, South-West meaning to the lower right, South meaning to the bottom, and South-East to the lower left. This is actually quite easy to forecast given the orientation of the movement of the Sun and of the Moon in the sky.

Concerning the discussion about the method that performed best, we note that the Muslim method was clearly the worse one, with all timing errors larger than 30 minutes. The Western method performed best for the magnitude and the duration, and did slightly better for the time of first contact, while the Datong method performed best for the times of the maximum and of the last contact. The competition between the Western and the Datong methods ended as 3-2 in football terms. Therefore it appears that Xu Guangqi succeeded in convincing the Emperor Chongzhen to entrust him with the reform of the calendar using the Western method, since this one performed considerably better for the important parameters of magnitude and duration. He probably also used in his favour previous eclipse predictions made with the Western method, such as the solar eclipse of 15 December 1610, which had been correctly predicted by Sabatino de Ursis, while it had not been foretold by Chinese astronomers (Udías, 2003: 40).

Table 1: Forecasts and observation of the eclipse of 21 June 1629, as reported in the Zhili yuanqi by Xu Guangqi, and comparison with the NASA reconstruction (in red). The forecasts which are closest to the NASA reconstruction are given in blue.

|  | First contact | Maximum | Last contact | Magnitude | Duration |
|---|---|---|---|---|---|
| Datong forecast (D) | $10^h 45^m$ | $11^h 45^m$ | $12^h 45^m$ | 0.340 | $2^h$ |
| Huihui forecast (H) | $11^h 45^m$ | $12^h 45^m$ | $13^h 45^m$ | 0.587 | $2^h$ |
| Western forecast (W) | $10^h 47^m$ | $11^h 36^m$ | $12^h 06^m$ | 0.2 | $1^h 19^m$ |
| Observation | $11^h 15^m$ |  | $12^h 45^m$ | 0.3 | $1^h 30^m$ |
| NASA corr. (N) | $11^h 01.3^m$ | $11^h 45.4^m$ | $12^h 30.0^m$ | 0.168 | $1^h 28.7^m$ |
| N-D | $16.3^m$ | $0.4^m$ | $-15.0^m$ | $-0.172$ | $-32.3^m$ |
| N-H | $-43.7^m$ | $-59.6^m$ | $-1^h 15.0^m$ | $-0.419$ | $-32.3^m$ |
| N-W | $14.3^m$ | $9.4^m$ | $24.0^m$ | $-0.032$ | $9.7^m$ |

A thorough collation of data concerning 120 eclipses recorded in Chinese sources from 3 August 134 BCE to 5 August 1785 has been conducted by Chen (1983). The source used by Chen for the eclipse of 21 June 1629 was in fact the *Lixue xiaobian* 曆學小辨 (Schall von Bell, 1631). In translating the timing in current notation, Chen (1983: 305) explained that he was considering the medium value for each unit of time; therefore, in the dual-hour and 100 *ke* system, the medium values were respectively 1 hour and 7.2. minutes. He assumed that Schall was still making use of this system in the *Lixue xiaobian,* so that the prediction of the time of first contact, indicated by Schall as *sizheng* 巳正 and 4 *ke* (Schall von Bell, 1631: f.2*v*), should correspond to $10^h 58.8^m$. This prediction is considerably more accurate than the one given by Xu Guangqi for the Western method, with a difference from the NASA reconstruction of only 2.5 minutes. However, since Chen (1983) reports about the prediction of the time of the first contact using the Western method, but gives no prediction for the other eclipse parameters nor with other methods, and since Adam Schall von Bell was not in Beijing at the time of the eclipse, we prefer to limit our analysis of ancient sources to the *Zhili yuanqi*.

Clearly the account of the eclipse given by Boorstin (1983) does not correspond to the report by Xu Guangqi, nor to the accurate NASA reconstruction. In particular the assumption that Jesuit astronomers had predicted that the eclipse would last only two minutes is clearly wrong, both because the eclipse lasted almost one hour and a half, and because the Jesuits had predicted a duration of one hour and 19 minutes. We suggest that the mistake might have been induced by a confusion of the magnitude of the eclipse, reported as 3 *fen* and 24 *miao*, meaning that a fraction of 0.340 of the Sun's diameter was obscured by the Moon at maximum, with a duration. This confusion seems to be present also in the report by Adam Schall von Bell (Chen, 1983).

We are puzzled by the difference of about 15 minutes between the observed eclipse times, as reported by Xu Guanqi, and those of the NASA reconstruction, since Xu Guanqi's timing accuracy should have been better, particularly for events close to the midday, although during the Ming Dynasty the development of mechanical clocks had stopped and more basic water clepsydras were in use (Steele and Stephenson, 1998). In fact the error seems to be a time shift, the actual eclipse duration having been recorded correctly within a minute or so. It is possible that this time shift could have been caused by a misalignment of the instruments available at the Beijing Ancient Observatory, as reported in a memorial of 1612 (Deane, 1994). In 1674 Ferdinand Verbiest made a new armillary sphere for the Beijing Observatory following the one made in Europe by Tycho Brahe in 1598 (Needham, 1959: 352). In the preface of his *Astronomiae apud Sinas Restitutae Mechanica* (see Golvers and Nicolaidis, 2009: 164), Verbiest writes that

> After the care of the Astronomical Bureau and the whole field of astronomy have been entrusted to me, some prominent people, who had attended our observations on the Astronomical Watchtower, had remarked that the Chinese astronomical instruments (at the Astronomical Bureau in Beijing) – although built after the model of the instruments of Guo Shoujing (responsible of the Astronomical Bureau more than 330 years ago i.e. during the time of the occupation of the Chinese Empire by the Mongols) – were in fact of a very primitive and clumsy construction when compared to my armillas and quadrants and sextants and other instruments by which I had recently observed the Heavens. They had established that these instruments were insufficient and not adapted, neither to our European astronomy, nor to their own and the Chinese Heavens (due to many errors, which I have indicated in my *Libri Organici*). Therefore, they petitioned the Emperor, and persuaded him to order me to construct new instruments; these were to be skillfully built according to the European type, and to be exposed on the Astronomical Watchtower as a perennial memorial to the Empire of the Eastern Manchu, after the ancient instruments had been removed from there. The Emperor immediately endorsed their proposal, and imposed the entire responsibility on my shoulders.

Therefore the instruments in use at the Astronomical Bureau at the time of the eclipse of 21 June 1629 were copies of instruments built for a different place, and were insufficient and not adapted to their function. It is quite possible that the one used for setting the time was misaligned by a few degrees causing a time shift by 15 minutes.

Finally, we have considered the possibility that the eclipse reports might have been changed since 1629, for example to satisfy the changing interests of those transmitting them (on this problem see e.g. Li et al., 2014). We have examined three different copies of the *Zhili yuanqi* 治曆緣起, by Xu Guangqi. One is at the Biblioteca Apostolica Vaticana (BAV, shelf mark: RGO III 233), the second one at the Archivum Romanum Societatis Jesu (ARSI, Jap.Sin II 15), and the third one is held at the Biblioteca Nazionale Centrale Vittorio Emanuele II (BNC, Fondo Gesuitico), also in Rome. We have carefully examined the pages corresponding to the eclipse of 21 June 1629 in these three editions and we have come to the conclusion that there is absolutely no difference in the text, although the quality of the paper used to print the volumes is not the same in the three editions. It is hard to establish with absolute certainty that the same woodblocks were used to print all of them, although this appears to be the case, even if they were probably printed at different times (a thorough analysis of the Ming versions of the *Zhili yuanqi* is in Chu, 2017). Therefore we may conclude that the reports of the important eclipse of 21 June 1629 have not been changed, at least for the three editions of the *Zhili yuanqi* that we have examined.

## 4 CONCLUSIONS

The eclipse prediction of 21 June 1629, a trivial episode as it may at first glance appear —one in which three contenders strove to prove the efficacy of their forecasting methods to win the trust of the Chinese court—was in fact a decisive event in determining the implementation of Western astronomy in China. About the larger errors made by Chinese and Muslim astronomers in the eclipse forecast we adopt the words of Terrentius, who pointed out that they proved the failure not of the astronomers, but of the methods they were following (Udías, 2003). In fact in 1281 the Chinese astronomer and engineer Guo Shoujing had estimated the length of the year as 365.2425 days (see Needham, 1959: 294), exactly the same as the value used for the Gregorian reform of the calendar in Europe three centuries later. The question about why this advanced astronomical knowledge in China did not correctly flow from the best astronomers to the practical applications such as the calendar, as it happened in Europe, is very interesting, but exceeds the purposes and limits of this paper. We simply suggest that the reasons may be connected with the larger independence and freedom of thoughts in Europe at the time.

The eclipse of 21 June 1629 was also decisive for the future of the Jesuit involvement with the Chinese Astronomical Bureau and the role the Jesuits played as mediators in the transfer of Western astronomical knowledge. This assumption is aptly demonstrated by the following declaration, made at the time and signed by ten officials serving at the Bureau:

> At first we also had our doubts about the astronomy from Europe when it was used in the *chi-ssu* year (1629), but after having read many clear explanations our doubts diminished by half, and finally by participating in precise observations of the stars, and of the positions of the sun and moon, our hesitations were altogether overcome. Recently we received the imperial order to study these sciences, and every day we have been discussing them with the Europeans. Truth must be sought not only in books, but in making actual experiments with instruments; it is not enough to listen with one's ears, one must also carry out manipulations with one's hands. All (the new astronomy) is then found to be exact. (Needham, 1959: 456).

Just as the events related to the eclipse of 21 June 1629 resulted from a fruitful encounter between people with different cultural backgrounds, we too, in the course of our work, have shared a very similar experience with several scholars world-wide, who have been very generous with their time and advice. It is our hope that such fruitful encounters will foster further research on the subject.

## 5 ACKNOWLEDGEMENTS

We would like to thank Christopher Cullen, Noël Golvers, Catherine Jami and Lü Lingfeng for useful discussions, an anonymous referee for valuablel comments that helped us improve our paper, and Shi Yunli for inviting us to a Conference on Science, Western Learning and Confucianism, held in Hefei in April 2019, where a first draft of this paper was presented and where we had many interesting discussions. The staff of BAV, ARSI and Biblioteca Nazionale in Rome have been generous with their time and expertise and we thank them for making available to us their copies of the *Zhili yuanqi*.

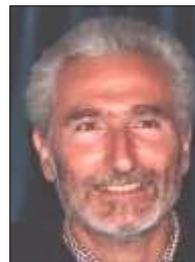

**Dr Sperello di Serego Alighieri** works at the Arcetri Astrophysical Observatory in Firenze, Italy. He became Doctor in Astronomy (summa cum laude) at University of Padova in 1975. After a fellowship from CNR he obtained a position as astronomer at the Padova Observatory in 1978, working on extragalactic astronomical research and on the development of astronomical instrumentation. From 1980 to 1984 he worked for the Astronomy Division of the Space Science Department of ESA at ESTEC in The Netherlands on the Faint Object Camera (FOC) for the Hubble Space Telescope (HST) and as Project Scientist for the Phase A study of the far UV space observatory Magellan. He built the prototype of the FOC, the instrument which granted the European participation in the HST project. From 1984 to 1990 he worked for ESA as HST Instrument Scientist at the Space Telescope European Coordinating Facility in Garching, Germany, helping European astronomers in the use of HST. Since 1990 he has been at the Arcetri Astrophysical Observatory in Firenze, Italy, working on astrophysical research in the extragalactic area (galaxy evolution, active galactic nuclei, observational cosmology). In the period 1995-1998 he was the Director of the Centro Galileo Galilei for the Italian telescope TNG on the Canary Island of La Palma. He is a keen traveler and recently he has been active on science communication and astronomy in Dante's Divina Commedia. See http://www.arcetri.astro.it/~sperello/

**Elisabetta Corsi** is Professor and Chair of Sinology and East Asian History at Sapienza University of Rome. She received her undergraduate training at Sapienza and carried out graduate studies at Beijing University, Hong Kong and Mumbai. She earned her PhD in Sinology at Sapienza. From February 2015 she has been Deputy Chair of the Humanities and Social Sciences Panel (Joint Research Scheme) of the Research Grants Council, Hong 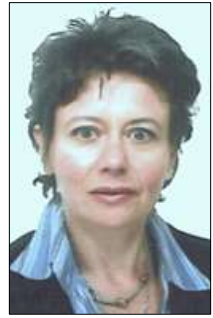 Kong. She is also a member of the Sapienza Quality Assurance Board and one of the founding members of the QuID workgroup for teaching enhancement. She is an invited Professor at the Pontifical Gregorian University. Her research interests focus on the role of Jesuit missionaries as scientific mediators during the early modern period, mainly in the field of 'mixed mathematics' – such as optics and linear perspective – and on the transmission of Aristotelian natural philosophy through the production, translation and circulation of printed works, mainly in Classical Chinese. She lectures on Chinese textual and exegetical traditions, reference tools and classification of knowledge in ancient China, as well as on material culture and connoisseurship in late imperial China, the relationship between technology and science, history of the book and printing, history of reading and writing, history of sinology and history of East Asia from a global perspective. She has published extensively in English, Spanish, Italian and Chinese. Her latest publication is "Idolatrous images and true images: European visual culture and its circulation in Early Modern China" in Del Valle, More and O'Toole's, *Iberian Empires and the Roots of Globalization* (2020, Vanderbilt University Press: 271-302.https://muse.jhu.edu/book/7295).